\documentstyle[epsf]{mn}
\input epsf.sty
\def\hmpc{~h$^{-1}$ Mpc~}
\def\hkpc{~h$^{-1}$ kpc~}

\def\eqw{$EQW({\rm [OII]})$}
\def\eqwmed{$\langle EQW({\rm [OII]}) \rangle$}
\def\sfrmed{$\langle SFR \rangle$}
\def\vmed{$\langle v \rangle$}

\title[A study of the core of the Shapley Concentration VI]
 { A study of the core of the Shapley Concentration: \\
 VI. Spectral properties of galaxies. 
\thanks{based on observations collected at the European Southern
Observatory, La Silla, Chile.} }
\author[ A.Baldi et al.]
{
Alessandro Baldi$^{1,2}$,
Sandro Bardelli$^{3}$,
Elena Zucca$^{3}$
\\ $^1$ Istituto di Fisica Cosmica G.Occhialini, 
via Bassini 15, I--20133 Milano, Italy
\\ $^2$ Dipartimento di Astronomia, Universit\`a degli Studi di Bologna,
via Ranzani 1, I--40127 Bologna, Italy
\\ $^3$ Osservatorio Astronomico di Bologna, 
via Ranzani 1, I--40127 Bologna, Italy
\\ E-mail: baldi@ifctr.mi.cnr.it, bardelli@bo.astro.it, zucca@bo.astro.it
}
\date{Received 00 - 00 - 0000; accepted 00 - 00 - 0000}
\begin{document}
\maketitle
\begin{abstract}

We present the results of a study of the spectral properties of galaxies in
the central part of the Shapley Concentration, covering an extremely wide
range of densities, from the rich cluster cores to the underlying supercluster
environment.
\\
Our sample is homogeneous, in a well defined magnitude range 
($17\le b_J \le 18.8$) and contains $\sim 1300$ spectra of galaxies at the 
same distance, covering an area of $\sim 26$ deg$^2$. 
These characteristics
allowed an accurate spectral classification, that we performed using a
Principal Components Analysis technique. 
\\
This spectral classification, together with the [OII] equivalent widths and the
star formation rates, has been used to study the properties of galaxies  
at different densities: cluster, intercluster (i.e. galaxies in the 
supercluster but outside clusters) and field environment.  
\\
No significant differences are present between samples at low density regimes 
(i.e. intercluster and field galaxies).
Cluster galaxies, instead, not only have values significantly different from 
the field ones, but also show a dependence on the local density.
\\
Moreover, a well defined morphology-density relation is present in the cluster 
complexes, although these structures are known to be involved in major merging 
events.
Also the mean equivalent width of [OII] shows a trend with the local 
environment, decreasing at increasing densities, even if it is probably
induced by the morphology-density relation.
\\
Finally we analyzed the mean star formation rate as a function of the density, 
finding again a decreasing trend (at $\sim 3\sigma$ significance level). 
Our analysis is consistent with the claim of Balogh et al. (1998) that the 
star formation in clusters is depressed.
\end{abstract}

\begin{keywords}
galaxies: distances and redshifts --
galaxies: spectra and morphology --
galaxies: clusters: general --
galaxies: clusters: individuals: A3528 - A3530 - A3532 - A3535 - A3556 -
                                 A3558 - A3562 --

\end{keywords}

%
\section{Introduction }

Since the discovery that galaxies in clusters are different from 
those in the field (Hubble \& Humason 1931), 
it has been recognized that the environment must play an important
role in determining the characteristics of the galaxy population.
In fact, at present the existence of a morphology-density
relation, spanning a large range of densities from rich clusters
(Dressler 1980) to groups (Postman \& Geller 1984), is generally accepted.
 
More intriguing is to determine whether the star formation 
is influenced by the environment or the different galaxy population in
clusters is merely a consequence of the morphology-density relation, and 
which mechanism is responsible for the change of the morphological mix.
 
From the observational point of view, other indications of differences 
between field and cluster galaxies come from the Butcher-Oemler effect
(i.e. the bluening of cluster galaxies as a function of redshift; 
Butcher \& Oemler 1984) and from the observation of ``anaemic" or ``HI 
deficient" spirals (i.e. with reduced content of HI with respect to field 
galaxies of the same morphological type; Bothun 1982) in local clusters. 
\\
Very recently, analysing the CNOC1 cluster sample in the redshift range
$0.2-0.5$, Balogh et al. (1998) found that the cluster environment 
not only affects the morphological mix but also suppresses the star formation.
\\
Various physical mechanisms (as for example galaxy harassment, tidal
stripping, ram pressure) have been proposed to explain the differences
between field and cluster galaxies: however, the influence of these 
phenomena on the changes in star formation rate remains still uncertain. 

In this context, an interesting point is that to compare properties of
field galaxies with those of galaxies which reside in rich superclusters,
but outside clusters.
\\ 
In fact, considering that up to scales of 2 \hmpc superclusters as the Great 
Wall are dominated by galaxy groups (Ramella, Geller \& Huchra 1992), 
a morphology-density relation may exist also in these lower density 
environments, leading to a difference in the morphological mix between
field and supercluster galaxies.
On the other hand, Hoffman, Lewis \& Salpeter (1995) found that the luminosity 
function of late type objects in the Great Wall does not differ from that
of field galaxies.

The aim of the present paper is to investigate the morphology-density 
relation in a wide range of densities, using an homogeneous sample of 
$\sim 1300$ spectra of galaxies in different environments, like rich
cluster cores, supercluster, field.
Our sample cover the central region of the Shapley Concentration, which 
has an average density excess in galaxies of 
$ \displaystyle{ {N} \over \bar{N} }= 11.3 \pm 0.4$ on a scale of 
$\sim 10$ \hmpc (Bardelli et al. 2000). 
This region is dynamically very active and is characterized by the presence 
of two high density complexes formed by merging clusters (Bardelli et al. 1994, 
1998a, 1998b, 2001). 
These complexes are embedded in a planar structure of 
$\displaystyle {{N} \over \bar{N}} \sim 4$, which resembles the 
Great Wall and represents the supercluster environment.
Finally, a subsample of galaxies from the ESP survey (Vettolani et al. 1997) 
is used to derive the reference values for the ``true" field environment 
$\displaystyle ({{N} \over \bar{N}} \sim 1)$. 

The paper is organized as follows. In Sect. 2 we present our classification
technique, based on the Principal Components Analysis, in Sect. 3 we 
describe our data and in Sect. 4 we show the results of the spectral
classification. In Sect. 5 we analyze the relations between the spectral 
morphology and the local environment and in Sect. 6 we present the results
about the emission line properties and the star formation rates. 
Finally, the results are discussed and summarized in Sect. 7. 
   
%
\section{Spectral classification }

Morphological classification of galaxies from optical images  
(Hubble 1936; de Vaucouleurs \& de Vaucouleurs 1961; Sandage 1975) is
generally performed analysing the luminosity profile of a galaxy. This method 
requires a very careful study of the images which implies a parametric fit
of the luminosity profile: in fact it is necessary to fit the luminosity 
profile of a galaxy to distinguish between a bulge and an exponential disk.
Moreover a galaxy may have a different appearance if observed using different 
filters or at different redshifts (O'Connell \& Marcum 1997).
\\ 
Even the advent of the Hubble Space Telescope does not prevent every existing 
method of morphological classification to be strongly dependent
on the image resolution, on the adopted filters and on the redshift of the 
galaxy.

An alternative approach to the morphological classification can be made if we 
consider the spectral energy distribution (SED), which resumes the main 
physical features of a galaxy. In fact, for a given galaxy, the SED measures 
the relative contribution of the various stellar populations and gives 
constraints to the amount of gas and to the mean metallicity.
\\
The spectral energy distribution is then useful to classify the galaxies
in a spectral sequence rather than in a morphological one. Generally, a galaxy 
is formed by three main components: gas, young stars and old stars. These 
components contribute to outline both the main morphological features (bulge, 
spiral arms, etc.) and the spectral ones (continuum shape, absorption and 
emission lines). Moreover, dealing with big amounts of data, spectra are 
easier to handle than bidimensional images, allowing an automatized 
classification of galaxies.
\\
In the past decade, such an automatization has been attempted by many authors. 
The most used technique, applied to surveys of galaxy spectra, is the Principal 
Components Analysis (hereafter PCA, Connolly et al. 1995; Folkes, Lahav \& 
Maddox 1996; Sodr\'e \& Cuevas 1997; Galaz \& de Lapparent 1998; Sodr\'e, 
Cuevas \& Capelato 1998; de Theije \& Katgert 1999). 
Using this technique, although with different 
approaches, these authors found that the whole distribution of spectral 
types can be described by only two parameters.
\\
Zaritsky, Zabludoff \& Willick (1995) assumed instead the a priori
hypothesis that each spectrum can be represented as a linear combination of 
three components: one representing old stellar populations (typical K star 
spectrum), one representing young stellar populations (typical A star spectrum)
and an emission lines component. Anyhow this approach has given results very 
similar to those obtained with the application of PCA.

\begin{figure*}
\centering
\leavevmode
\epsfxsize=0.49\hsize \epsfbox{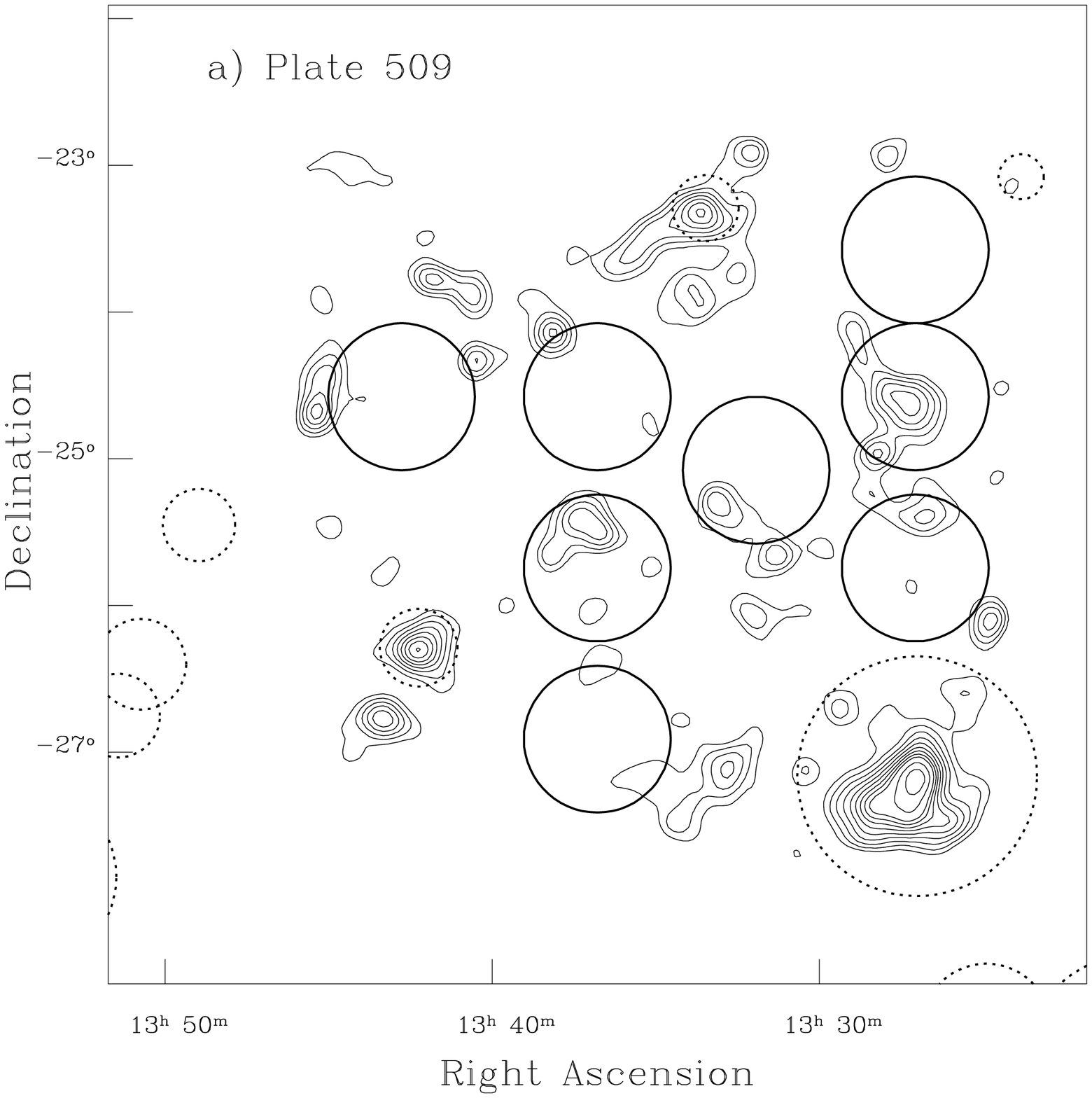} \hfil
\epsfxsize=0.49\hsize \epsfbox{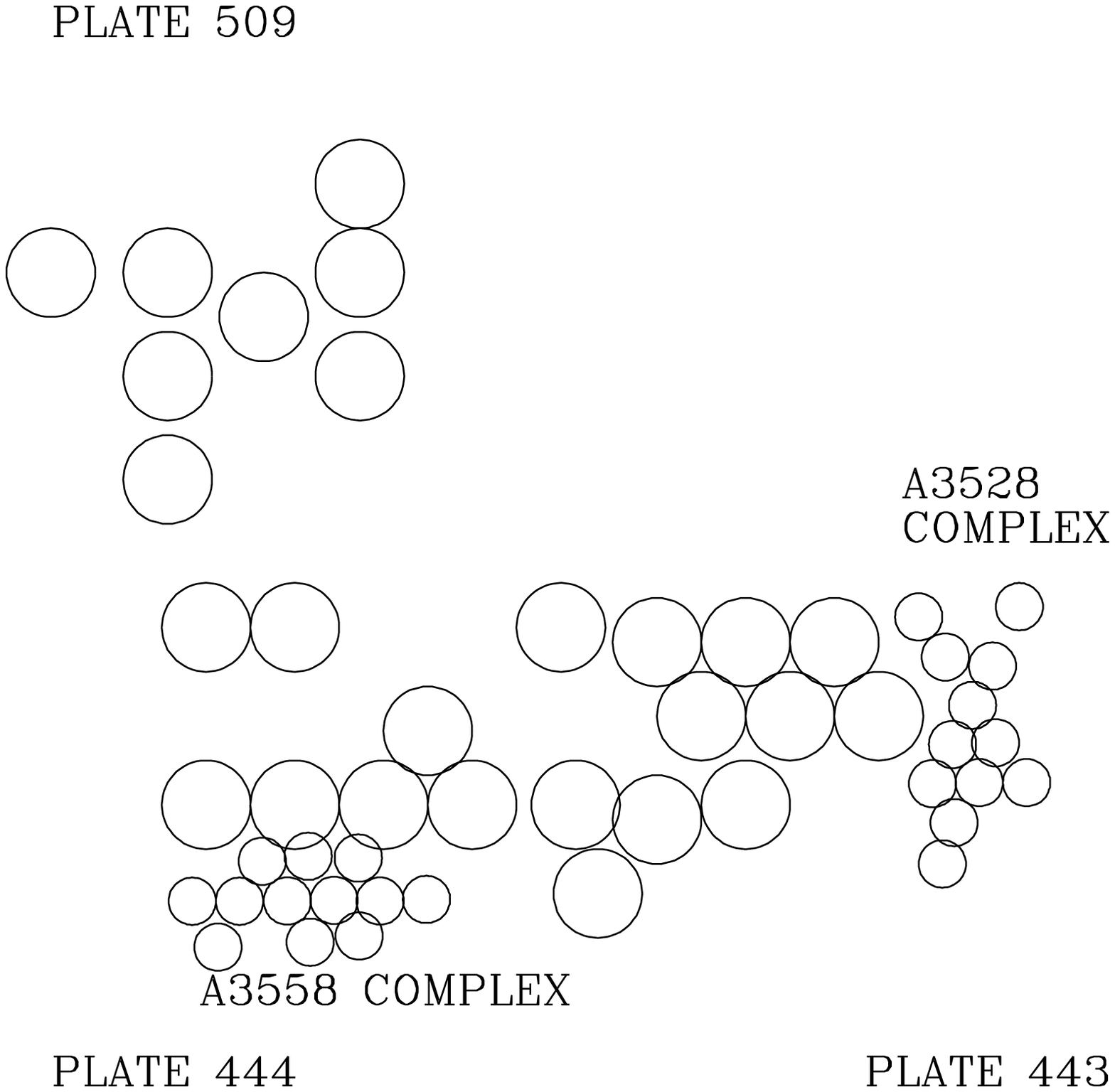} \hfil
\centering
\leavevmode
\epsfxsize=0.49\hsize \epsfbox{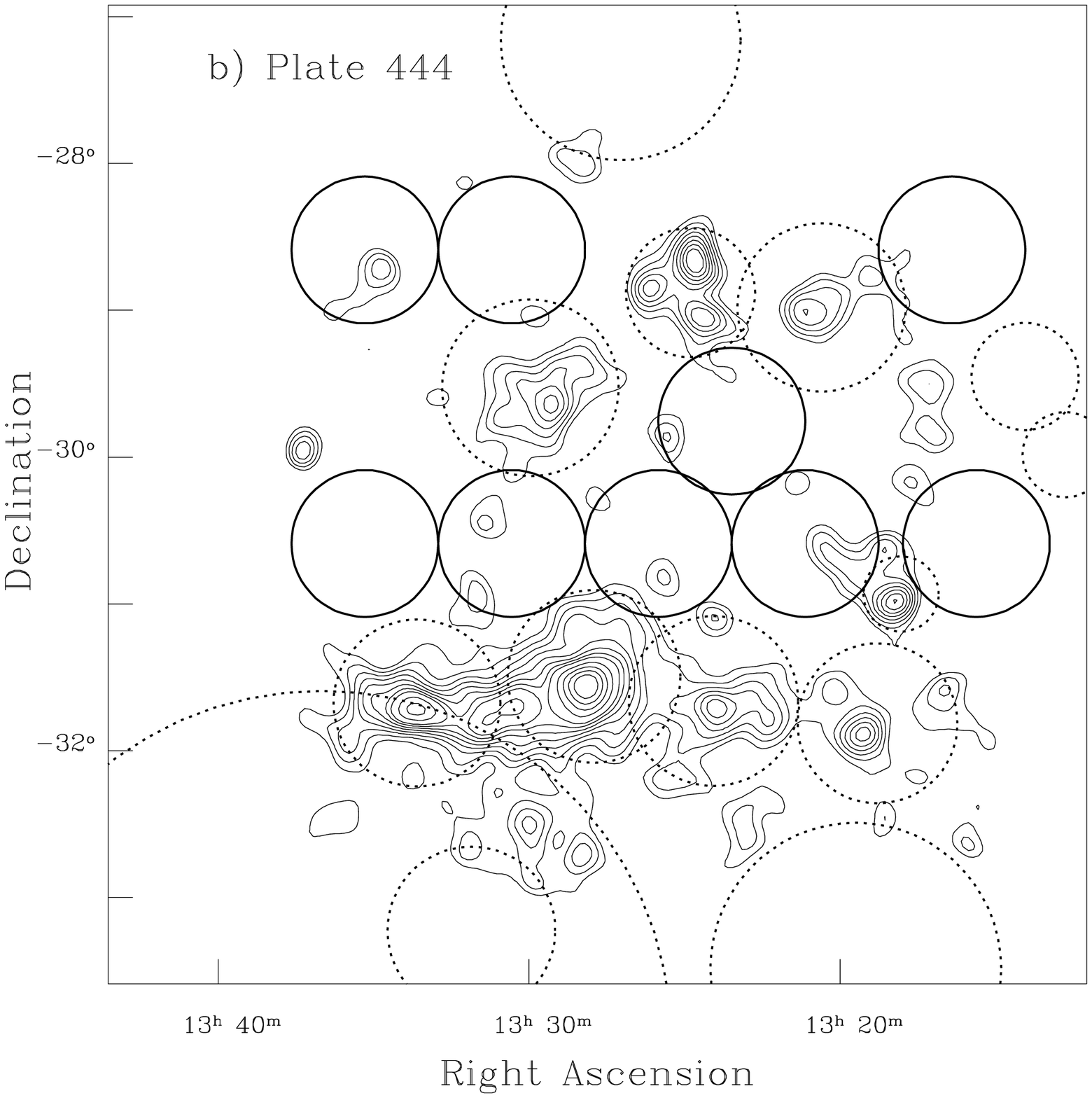} \hfil
\epsfxsize=0.49\hsize \epsfbox{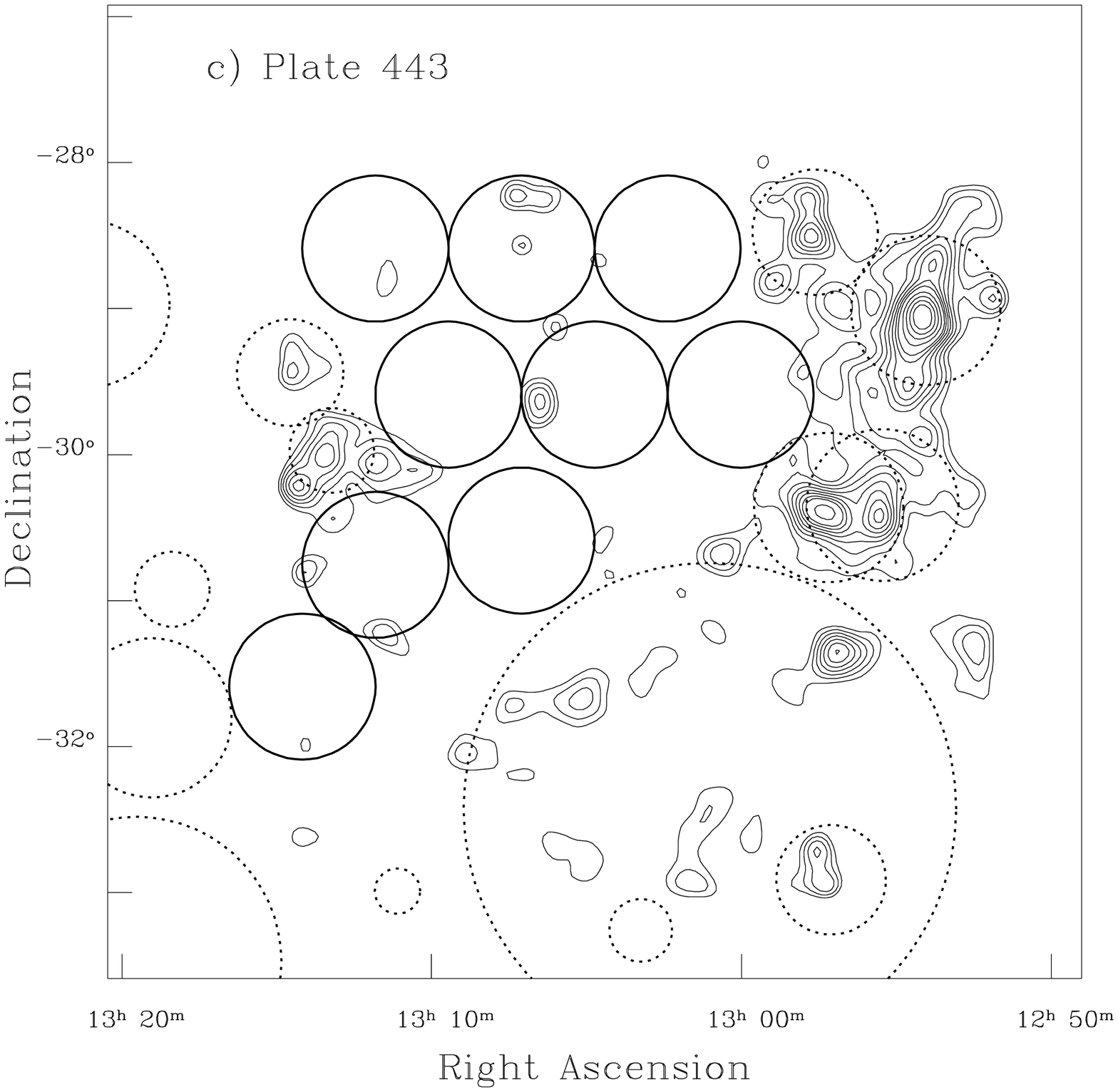} \hfil
\caption[]{Isodensity contours of the bidimensional distribution of the 
galaxies in the $b_J$ magnitude bin $17-19.5$ in the UKSTJ plates which
cover the central part of the Shapley Concentration.
The data have been binned in $2$ $\times$ $2$ arcmin 
pixels and then smoothed with a Gaussian with a FWHM of $6$ arcmin. For the
Abell clusters present in the plates, circles of one Abell radius have been
drawn (dashed curves). Solid circles represent the MEFOS fields.
\\
a) Plate 509; b) Plate 444; c) Plate 443. The two most evident systems are 
the A3558 (plate 444) and the A3528 (plate 443) cluster complexes. 
The relative positions of the fields is shown in the upper right panel, where
small circles represent OPTOPUS observations on the cluster complexes
(which for clearness were not drawn on the isodensity contours).}
\label{plates}
\end{figure*}

%
\subsection{The PCA technique \label{pcatech}}

The method we have chosen to perform the spectral classification of our galaxy 
sample is the Principal Components Analysis. An exhaustive description of 
the PCA technique can be found in Murtagh \& Heck (1987) or in Kendall (1980):  
here we underline only its main features of interest for our work.
\\
The PCA applies to a sample of $N$ objects (in our case $N$ spectra) which can 
be described by $M$ coordinates (in our case, for each spectrum, the fluxes of 
single pixels). In this $M$-dimensional hyperspace, each object is represented 
by a point and the sample is a cloud of points. The main idea of the PCA is 
to find an orthonormal base of eigenvectors (of dimension $P < \min [N,M]$) 
whose linear combination represents the sample. The eigenvectors are found 
by minimizing the euclidean distance of the points from the new base:  
these new axes are called Principal Components (PCs). 
The eigenvalues give the variance of each PC. An higher 
value of the variance indicates a lower distance between the cloud of points 
and the new axis relative to the eigenvalue, then a better description of the 
sample. 
\\
Each spectrum {\bf S} is normalized by its norm as:
\begin{equation}
{\bf S}^{norm}=\frac{\bf S}{\sqrt{\sum_{j=1}^M {\bf S}_j^2}}
\end{equation}
Other methods of normalization can be used (for instance a flux normalization
at a given wavelength), but Connolly et al. (1995) have demonstrated that
the various methods of normalization do not have relevant effects onto the
PCA results.
\\ 
The techniques generally used to reduce the input matrix are three:
the sum of squares and cross product matrix method (SSCP), the 
variance-covariance matrix method (VC) and the correlation matrix method (C).
The first method (SSCP) does not apply any rescaling on the data nor 
centre the cloud of points. The normalized spectra then lie on the
surface of an $M$-dimensional hypersphere of radius 1, and the first PC has 
the same direction of the average spectrum. The second method (VC) places 
instead the new origin onto the centroid of the sample. The latter method (C) 
centres the cloud of points and rescales the data in such a way that the 
distance between variables is directly proportional to the correlation between 
them.
\\
It is worth to stress that neither the PCs nor the projections on them obtained
with the three different methods are the same. However, if the  
cloud of points is concentrated in a small portion of the hypersphere of 
radius 1, then the first PC obtained with VC method will have almost the same
direction of the second PC obtained from SSCP method (Galaz \& de Lapparent 
1998).
\\
After performing the PCA, it becomes possible to reconstruct each spectrum
{\bf S}$^{norm}$ as:
\begin{equation}
{\bf S}_{approx}=\sum_{k=1}^{N_{PC}} \alpha_k {\bf E}_k
\end{equation}
where {\bf S}$_{approx}$ is the {\bf S}$^{norm}$ spectrum reconstructed using
only the first $N_{PC}$ PCs and $\alpha_k$ is the projection
of {\bf S}$^{norm}$ onto the eigenspectrum (principal component) {\bf E}$_k$.

%
\subsection{Application of the PCA to the data \label{applic}}

The three methods described in the previous section have been tested on our 
dataset. We decided to use the SSCP method for the following reasons.
First, the first three PCs are sufficient to reconstruct $\sim$99\% in flux 
of the sample, i.e. at the same percentage of reconstructed flux, a lower 
number of components is enough. As a comparison, the VC method gives a 
percentage of $\sim$87\% and the C method gives a percentage less than 80\% 
for the first three principal components. Moreover, plotting $\alpha_1$ vs. 
$\alpha_2$, which are the projections onto the first two PCs, the diagram has 
a smaller scatter, indicating a better definition of a spectral sequence 
(see below). Finally, the first three principal components have a 
straightforward physical interpretation, although this is not an objective 
criterion for choosing a method.
\\
Using the SSCP method it is then possible to perform a spectral classification
dependent only on three parameters. Moreover, following Galaz \& 
de Lapparent (1998), we can apply a change of coordinates, from ($\alpha_1$, 
$\alpha_2$, $\alpha_3$), the projections onto the PCs, to ($r$, $\delta$, 
$\theta$), the radius, the azimuth and the polar angle measured from the 
equator, respectively:
\begin{eqnarray}
\alpha_1&=&{\it r} \cos \theta \cos \delta\\
\alpha_2&=&{\it r} \cos \theta \sin \delta\\
\alpha_3&=&{\it r} \sin \theta
\end{eqnarray}
It is then possible to isolate the values of $\delta$ and $\theta$ 
from $r$ (which is equal to 1, because of the applied normalization):
\begin{eqnarray}
\delta&=&\arctan \left(\frac{\alpha_2}{\alpha_1}\right)\\
\theta&=&\arctan \left\{\left(\frac{\alpha_3}{\alpha_2}\right) \sin 
\left[\arctan \left(\frac{\alpha_2}{\alpha_1}\right)\right]\right\}
\end{eqnarray}
The parameters used to perform the spectral classification are then reduced to 
only two: we have in fact eliminated the magnitude parameter (represented by
$r$) having performed a normalization of the spectra. 
In general, morphological types do depend on the
magnitude, because of the differences in luminosity functions of galaxies
going from ellipticals to irregulars (see f.i. Binggeli, Sandage \& Tammann 
1988); however, as described below, we have a very small range in magnitudes 
that allows us to neglect this parameter.
\\
Moreover it is possible to associate a physical meaning to $\delta$ and 
$\theta$. In fact, $\delta$ represents the contribution of the blue over the 
red part of the spectrum, i.e. of the young stellar population relative to the 
older one, while $\theta$ represents the importance of emission lines in a 
galaxy spectrum and could then be an indicator of star formation (see below).

\begin{table*}
\caption{\label{compl} Number of objects in the various samples. }
\begin{tabular}{lcccc}
\hline
 & $N_{tot}$ & $N_{z}$ & $N_{Shapley}$ & $N_{spectra}$ \\
\hline
A3528 complex $(17\le b_J \le 18.8)$ & 674 & 504 & 402 & 350 \\
A3558 complex $(17\le b_J \le 18.8)$ & 722 & 464 & 421 & 264 \\
Intercluster sample $(17\le b_J \le 18.8)$ & 1839 & 459 & 166 & 161 \\
\hline
A3528 complex $(b_J \le 19.5)$ & 1439 & 678 & 527 & 447 \\
A3558 complex $(b_J \le 19.5)$ & 1582 & 727 & 633 & 397 \\
\hline
\end{tabular}
\end{table*}

%
\section{The data }
\label{thedata}

The photometric data catalogue is the COSMOS/UKST galaxy catalogue of the 
southern sky (Yentis et al. 1992) obtained from automated scans of UKST $J$ 
plates by the COSMOS machine.
\\
Spectroscopic observations were obtained using the ESO 3.6-m telescope in La
Silla equipped with the OPTOPUS multifibre spectrograph (Lund 1986), for 
observations in the regions of cluster complexes, and with the MEFOS multifibre 
spectrograph (Bellenger et al. 1991; Felenbok et al. 1997) for observations in 
the intercluster region. In Figure \ref{plates} we show the position of the
observed OPTOPUS and MEFOS fields, superimposed on the galaxy isodensity
contours.
\\
A further spectral sample from the ESO Slice Project (ESP) galaxy redshift 
survey (Vettolani et al. 1997) has been considered in order to have a 
comparison of supercluster spectra with those of ``true" field galaxies. 
\\
From all these samples, we have considered only galaxies in the velocity 
range $10000 - 22500$ km/s, in order to limit the spectral analysis to the 
physical extension of the Shapley Concentration (Bardelli et al. 2000). 
Moreover, given the fact that the intercluster survey is limited to 
the magnitude range 17 $\le$ $b_J$ $\le$ 18.8, when we compare the cluster 
galaxy properties with the intercluster ones, we use subsamples with these
magnitude limits. This choice avoids possible biases deriving from sampling of 
luminosity functions of various morphological types at different magnitudes.
On the contrary, when we consider only properties of cluster galaxies, 
we use the samples in the whole magnitude range ($b_J \le 19.5$).

In Table~\ref{compl} the relevant numbers for each sample are given. 
In column (1) we give the sample name, in column (2) the number of 
COSMOS/UKST galaxies inside the survey area and in column (3) the number 
of galaxies with redshift (including literature measurements).
Column (4) reports the number of objects belonging to the Shapley Concentration:
among these, we could classify only objects with spectra observed by us.
This number is reported in column (5).

%
\subsection{The A3528 complex }

The region considered to sample the A3528 complex is part of 
the UKST $J$ plate 443 and has a dimension on the plane of the sky 
of $\alpha \times \delta \sim 2\fdg7 \times 3\fdg8$, in the range 
$12^h 51^m < \alpha(2000) <13^h 15^m $ and 
$-31^o 36\arcmin < \delta(2000) < -28^o 00\arcmin$ (Bardelli, 
Zucca \& Baldi 2001).
\\
The spectroscopic observations were performed with the OPTOPUS multifibre 
spectrograph; we used the ESO grating $\#\ 15$ allowing a resolution of 
$\sim 12$ \AA\ in the wavelength range $3700 - 6024$ \AA. The detectors were 
Tektronic $512\times 512$ CB CCDs (ESO $\# 16$ for 1991 run; ESO $\# 32$ for 
1993 run) with a pixel size of $4.5$ \AA. Detector $\# 32$ has a particularly 
good responsive quantum function in the blue ($\sim 70\%$ at $4000$ \AA), if 
compared with detector $\# 16$.
\\
All the details on the reduction steps can be found in Bardelli et al. (1994):
however, it could be important to stress that we normalized the fiber 
transmission rescaling the spectra in order to have the same 
continuum--subtracted flux of the sky emission line [OI]$\lambda$5577.

%
\subsection{The A3558 complex }

The region considered to sample the A3558 complex lies on the
UKST $J$ plate 444 and has a dimension on the plane of the sky 
of $\alpha \times \delta \sim 3\fdg2 \times 1\fdg4$, in the range 
$13^h 22^m 06^s < \alpha(2000) <13^h 37^m 15^s$ and
$-32^o 22\arcmin 40\arcsec < \delta(2000) < -30^o 59\arcmin 30\arcsec$
(Bardelli et al. 1994, Bardelli et al. 1998a).
\\
The spectroscopic observations were performed with the same instruments and
the setup used for the A3528 complex survey (see above).  
It is worth noting from Table~\ref{compl} that the number of spectra 
available for the classification in this sample is lower than in the
sample relative to the A3528 complex: this is due to the large number of 
galaxies with a redshift determination in the literature for the A3558 complex
and then not re-observed in our survey.

%
\subsection{The intercluster sample}

The intercluster survey, described and analyzed in detail in Bardelli et al.
(2000), has been performed in order to study the distribution
of galaxies in the Shapley Concentration outside the clusters.
The spectroscopic observations covered part of the UKST $J$ 
plates 443, 444 and 509 and  were obtained with the MEFOS spectrograph.
 In order to maximize the performances of this instrument, we adopted
the magnitude range $17-18.8$.  
\\
The spectra, obtained with the CCD TEK512 CB \# 32 and the ESO grating $\# 15$,
have a resolution of $\sim 12$ \AA~and a pixel size of $\sim 4.6$ \AA: 
data were reduced and cross-correlated in the same way as done for the 
cluster complexes.

%
\subsection{The ESP sample \label{ESPsample}}

The ESP galaxy redshift survey (Vettolani et al. 1997) extends over a strip of 
$\alpha\times\delta$=$22^\circ\times 1^\circ$, plus a nearby area of 
$5^\circ\times 1^\circ$, five degrees west of the main strip, in the South 
Galactic Pole region. The whole ESP sample, together with the detailed
description of the survey, is available in Vettolani et al. (1998). 
\\
To our aims it is sufficient to remember 
that the spectroscopic observations were performed at the ESO 3.6-m telescope 
in La Silla, using the multifibre spectrographs OPTOPUS and MEFOS with the
same instrumental setup used for our surveys on the Shapley Concentration.
However, given the fact that the wavelength coverage of some ESP spectra
is limited to 3900 - 6100{\AA}, in order to include the [OII]$\lambda$3727
line in all the rest frame spectra we are forced to use only 
galaxies with $v>15800$ km/s, instead $v>10000$ km/s as done for all the other 
samples. Applying also the $17-18.8$ magnitude cut, we obtained a sample
of $\sim 300$ galaxies. The mean galaxy density in this sample is
consistent with that of the whole ESP survey.    
\\
The ESP galaxies are supposed to be ``true" field galaxies and should 
represent a well defined reference sample to analyse the variations in 
morphological mixing inside the Shapley supercluster.

\begin{figure}
\epsfxsize=\hsize
\epsfbox{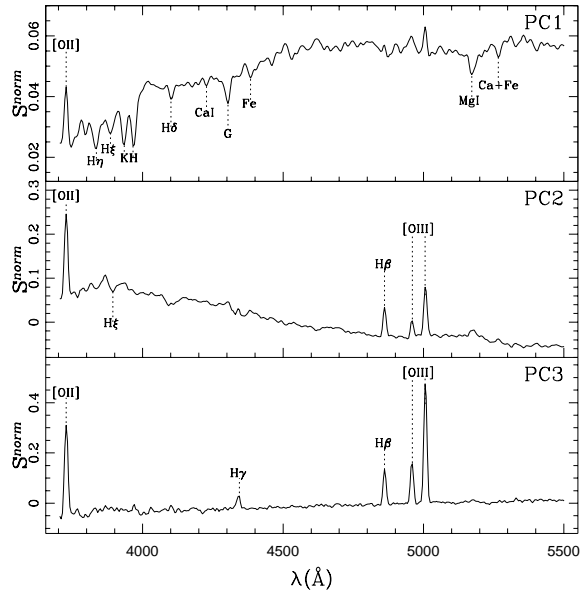}
\caption[]{The first three principal components (PCs) obtained by PCA. The main
spectral lines of each PC are represented.}
\label{PCs}
\end{figure}

\begin{table}
\caption{\label{kenn}Galaxies of Kennicutt's sample.}
\begin{tabular}{|ll|ll|}\hline

name&type&name&type\\
\hline
NGC 3379&E0&NGC 3147&Sb\\
NGC 4472 (1)&E1/S0&NGC 3227 (3,4)&Sb\\
NGC 4648&E3&NGC 3627&Sb\\
NGC 4889 (2)&E4&NGC 5248&Sbc\\
NGC 3245&S0&NGC 6217&SBbc\\
NGC 3941&SB0/a&NGC 2276 (3)&Sc\\
NGC 4262 (1)&SB0&NGC 2903&Sc\\
NGC 5866&S0&NGC 4631&Sc\\
NGC 1357&Sa&NGC 4775&Sc\\
NGC 2775&Sa&NGC 6181&Sc\\
NGC 3368&Sab&NGC 6643&Sc\\
NGC 3623&Sa&NGC 4449&Sm/Im\\
NGC 1832&SBb&NGC 4485 (3)&Sm/Im\\
\hline
\end{tabular}

\medskip
(1) Virgo cluster member;
(2) Coma cluster member; (3) strongly interacting or merging galaxy; (4)
galaxy with a Seyfert 2 nucleus.
\end{table}

%
\subsection{Preparing the data for the PCA algorithm }

The spectral sample described in previous sections has been prepared for
the application of the PCA via the SSCP method. 
The spectra have been de-redshifted to the rest frame, reduced at the same
rest frame dispersion (4.5 {\AA}/pix) and cut in the same wavelength range
(3705 - 5500 {\AA}). This wavelength band has been chosen in order to observe 
all the spectral features between the [OII] $\lambda$3727 emission line and
the Ca+Fe $\lambda$5269 absorption line. The number of pixels (i.e. the number 
of coordinates describing each spectrum in the PCA) is 400. 
\\
In order to have a link between our PCA spectral classification and the 
classical morphologies of galaxies, we added to our sample a set  
of spectra of nearby galaxies from the spectrophotometric atlas of Kennicutt 
(1992a): these spectra are of good quality and refer to galaxies with
known morphology. From the original atlas, we used the 26 normal galaxies 
listed in Table~\ref{kenn}.
These spectra have been reduced to the same wavelength range and dispersion
of our sample: indeed, it is necessary to run the PCA procedure at the same 
time for all the spectra (our data plus Kennicutt's data)
in order to obtain a relative scale to define the spectral sequence.

%
\section{Results from the PCA }
\label{results}

\begin{figure*}
\centering
\leavevmode
\epsfxsize=0.8\hsize
\epsfbox{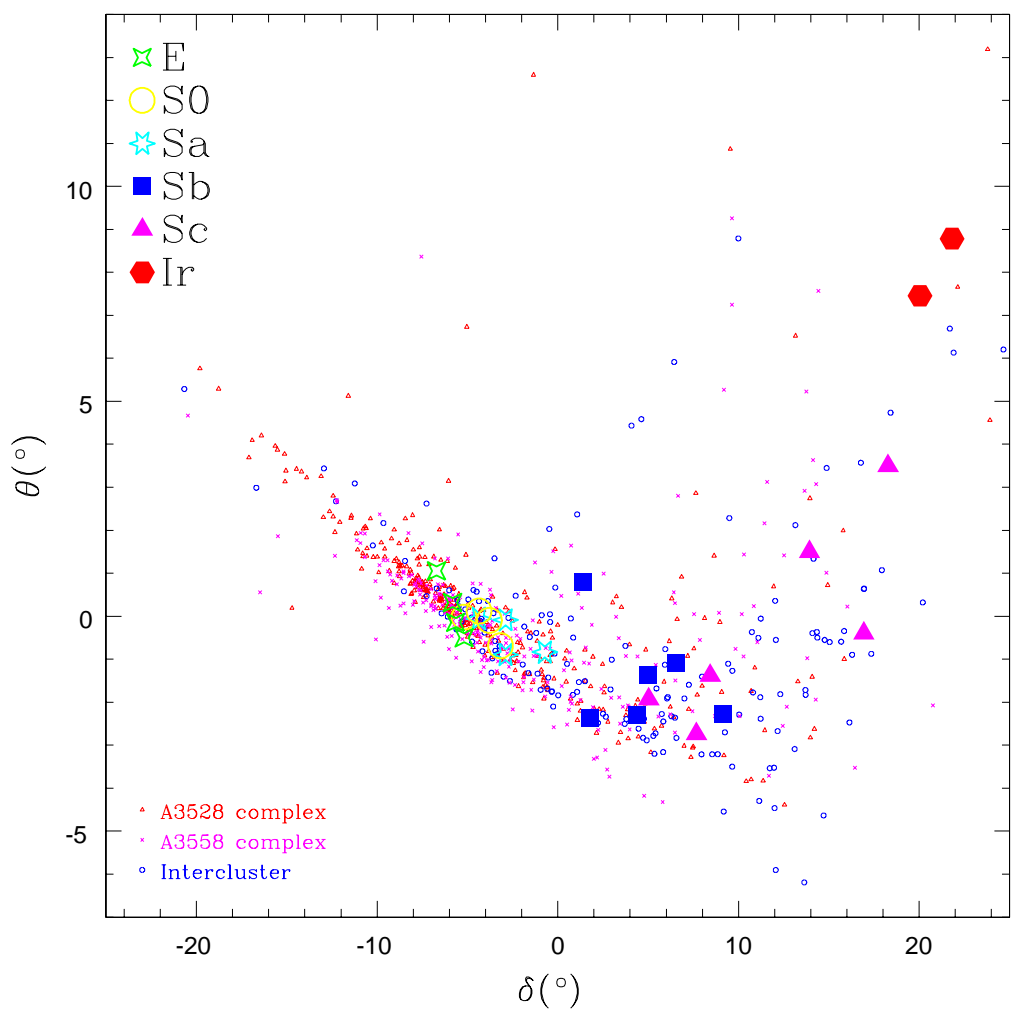} \hfil
\caption[]{The classification diagram $\delta$-$\theta$ for the spectral sample
in the Shapley Concentration plus the Kennicutt's sample (big symbols). 
There are a few objects with values of $\delta$ and/or $\theta$ which lie 
outside the range represented in this figure: they have spectra with very 
strong emission lines or with very poor signal-to-noise ratio. } 
\label{delthe}
\end{figure*}

The PCA classification has been applied to the sample of spectra 
from the cluster complexes, the intercluster field, the ESP survey subsample 
and the Kennicutt's atlas, in the magnitude range $17\le b_J \le 18.8$. 
In order to make a relative comparison between the morphological mix of
the two cluster complexes, we ran the PCA also in the whole magnitude range
($b_J \le 19.5$) for these samples, finding similar results.
\\
The first three principal components are represented in Figure \ref{PCs}.
The first principal component (PC1) contributes to 96.7\% in flux of the 
sample and is characterized by a prominent red continuum relative to the blue 
one. It shows spectral features like a clear 4000{\AA} break, a number of 
absorption lines like the pair K $\lambda$3934 and H $\lambda$3969, 
H$\delta$ $\lambda$4102, G $\lambda$4304 and MgI $\lambda$5175, and the 
forbidden emission line of [OII] $\lambda$3727. 
The continuum of this spectrum resembles closely that of an early-type galaxy 
like an E or an S0: however, the presence of the [OII] emission line indicates
that this PC has also some characteristics of early spirals, like Sa and Sb. 
The contribution of PC1 is dominant because our sample is mainly composed by
early-type galaxies. 
\\
The second principal component (PC2) contributes on average to 2.0\% in flux
of the sample. The blue part of the spectrum is enhanced with respect to the 
red one and the [OII] emission line is clearly more prominent than in PC1;
moreover, the emission features of H$\beta$ $\lambda$4861 and of [OIII] 
$\lambda\lambda$4959,5007 clearly appear. This eigenspectrum is 
characterized by less important absorption features and 
resembles the spectrum of a galaxy with star formation activity, where the 
dominant population is constituted by the blue young stars, which dominates the 
continuum. The ongoing star formation activity in such a type of galaxy can be 
detected by the presence of nebular emission lines too.
\\
The third principal component (PC3) contributes on average to 0.3\% in flux
of the spectral sample. Its continuum is almost flat and dominated by the
emission lines, stronger than in PC2, of [OII], H$\beta$ and [OIII]: it 
presents also the H$\gamma$ $\lambda$4341 emission line, although not so 
prominent like the others, which is not present in PC2. 
Hence this eigenspectrum represents essentially the emission features in a 
galaxy spectrum, which can be related, expecially in the case of [OII], to the
presence of an ongoing star formation activity (Kennicutt 1992b).
\\
The other principal components contribute marginally to the flux of our
spectral sample (0.12\% for the fourth, 0.07\% for the fifth, etc.) and have
not a straightforward physical interpretation. 
Therefore we have decided to neglect them and to perform the spectral 
classification using only the first three eigenspectra, which contribute 
together to $\sim$99\% in flux of the sample.
\\
Then, as described in Sect. \ref{applic}, we applied a change of coordinates 
from ($\alpha_1$, $\alpha_2$, $\alpha_3$) to ($\delta$, $\theta$):
the $\delta$-$\theta$ plot is shown in Figure \ref{delthe}, for all the
spectra of the sample in the Shapley Concentration. 
The Kennicutt's galaxies are plotted with big points and with different symbols 
for each Hubble type. To simplify the figure, we have decided to put barred 
and normal spirals (like Sa and SBa) in the same category and intermediate 
types have been put in the earliest category (for instance, an Sab galaxy has 
been marked as Sa). The two galaxies labelled Ir (irregulars) are Sm/Im.
\\
From this figure, it is clearly visible that the galaxies follow a
well defined sequence and that the Kennicutt's objects are located in
a succession of increasing values of $\delta$, going from the ellipticals to 
the irregulars. For this reason, in the following we will use $\delta$ to
classify galaxies. A similar result has been found by Sodr\'e
\& Cuevas (1997) which, applying the PCA (although not with SSCP method) to 
the entire Kennicutt's sample, found that, for normal galaxies, only one 
parameter is necessary to discriminate among morphological types along the 
Hubble sequence.
\\
This behaviour can be understood remembering that $\delta$
represents the importance of the blue part of the continuum with respect to 
the red one. The $\theta$ parameter is an indicator of the emission line
strength in a galaxy spectrum: this fact explains the increasing spread in 
$\theta$ moving along the sequence toward late types. 
\\
The few objects which strongly deviate from the sequence 
have mostly very low signal-to-noise spectra, except for a galaxy in the A3528 
complex (located at $\delta\sim -5^\circ$ and $\theta\sim 7^\circ$) which 
has a spectrum characterized by a red (elliptical-like) continuum
and by the contemporary unusual presence of very strong [OIII] emission lines.
\\
It is worth noting that the separation in $\delta$ between 
two different Hubble types is smaller for earlier types than for later 
ones: in fact , as shown by Kennicutt (1992a, 1992b), the difference between 
the spectrum of an E/S0 and an Sb is very subtle and consists only in a 
slight decreasing of the 4000{\AA} break importance and in a progressive 
growth of the H$\alpha$+[NII] equivalent width, which often reflects in a 
growth of the [OII] equivalent width. The differences become more important 
starting from Sc spirals.
\\ 
However it is important to note that this ``compression" in the early type 
region of the sequence may be due, at least in part, to the fibre angular 
dimension. In fact both OPTOPUS and MEFOS fibres have a dimension projected on 
the plane of the sky of $\sim2.5\arcsec$. This means that, at the average 
distance of the Shapley Concentration ($z\sim0.05$), they cover only the 
central $\sim$2\hkpc of a galaxy, i.e. for early spirals they can observe only 
their bulge region, where the stellar population and the gas component are 
more similar to those of E/S0 galaxies than in the spiral arms. Hence this may 
result in a ``compression" of the spectral sequence until we reach the 
intermediate spiral region, where a smaller bulge extension allows to observe 
part of the spiral arms too.

%
\subsection{Relation between the $\delta$ parameter and the galaxy colours}

\begin{figure}
\epsfxsize=\hsize
\epsfbox{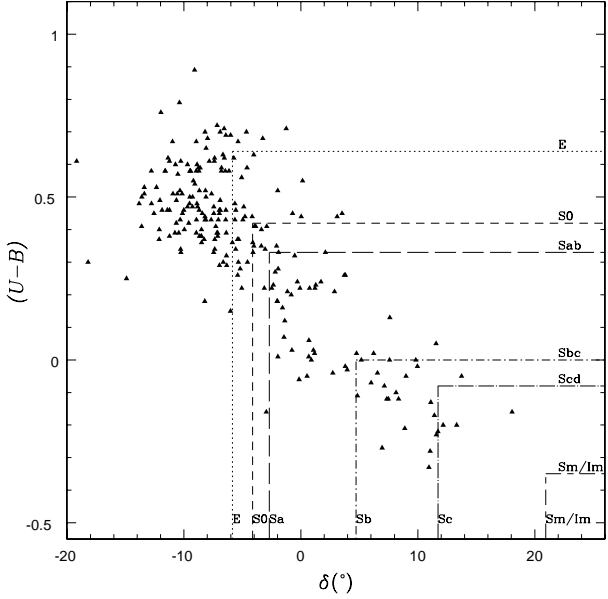}
\caption[]{Relation between $\delta$ and $(U-B)$ colours. 
See the text for details on the lines. }
\label{coldelta}
\end{figure}

Metcalfe, Godwin \& Peach (1994) give photometric colours for many galaxies in 
the A3558 complex. Hence, we have decided to compare our classification 
parameter $\delta$ with galaxy colours, in order to possibly extrapolate a 
value of $\delta$ for the galaxies not observed in our spectroscopic survey.
\\ 
In the magnitude range 17 $\le$ $b_J$ $\le$ 18.8, there are 230 galaxies with 
both a $(U-B)$ colour and a spectral 
classification: they are plotted with solid triangles in Figure \ref{coldelta} 
in a diagram $\delta$ vs. $(U-B)$. 
As expected, a relation between the two quantities is clearly evident: lower 
values of $(U-B)$ correspond to higher values of $\delta$.
The vertical lines in this figure represent the mean value of $\delta$  
corresponding to different Hubble types of the Kennicutt's galaxies, while the 
horizontal lines correspond to the mean $(U-B)$ colours expected for each 
Hubble type from Fukugita, Shimasaku \& Ichikawa (1995). 
Note that Fukugita et al. (1995) do not give a mean colour for 
each distinct class of spirals, as we do for the Kennicutt's galaxies, 
but colours for intermediate classes of spirals from Sab to Scd. 
\\
The existence of such a relation allows us to use the photometric data from 
Metcalfe et al. (1994) to roughly classify galaxies in the A3558 complex
with redshift from the literature but without an available spectrum. 
This addition has become necessary in order to have a more uniform coverage, 
especially in the central parts of A3558, where the majority of literature
data are located, and results in an increase of $\sim 45\%$ of the sample 
size. 

%
\section{Spectral morphology vs. environment  }

%
\subsection{Global properties }

\begin{figure}
\epsfxsize=\hsize 
\epsfbox{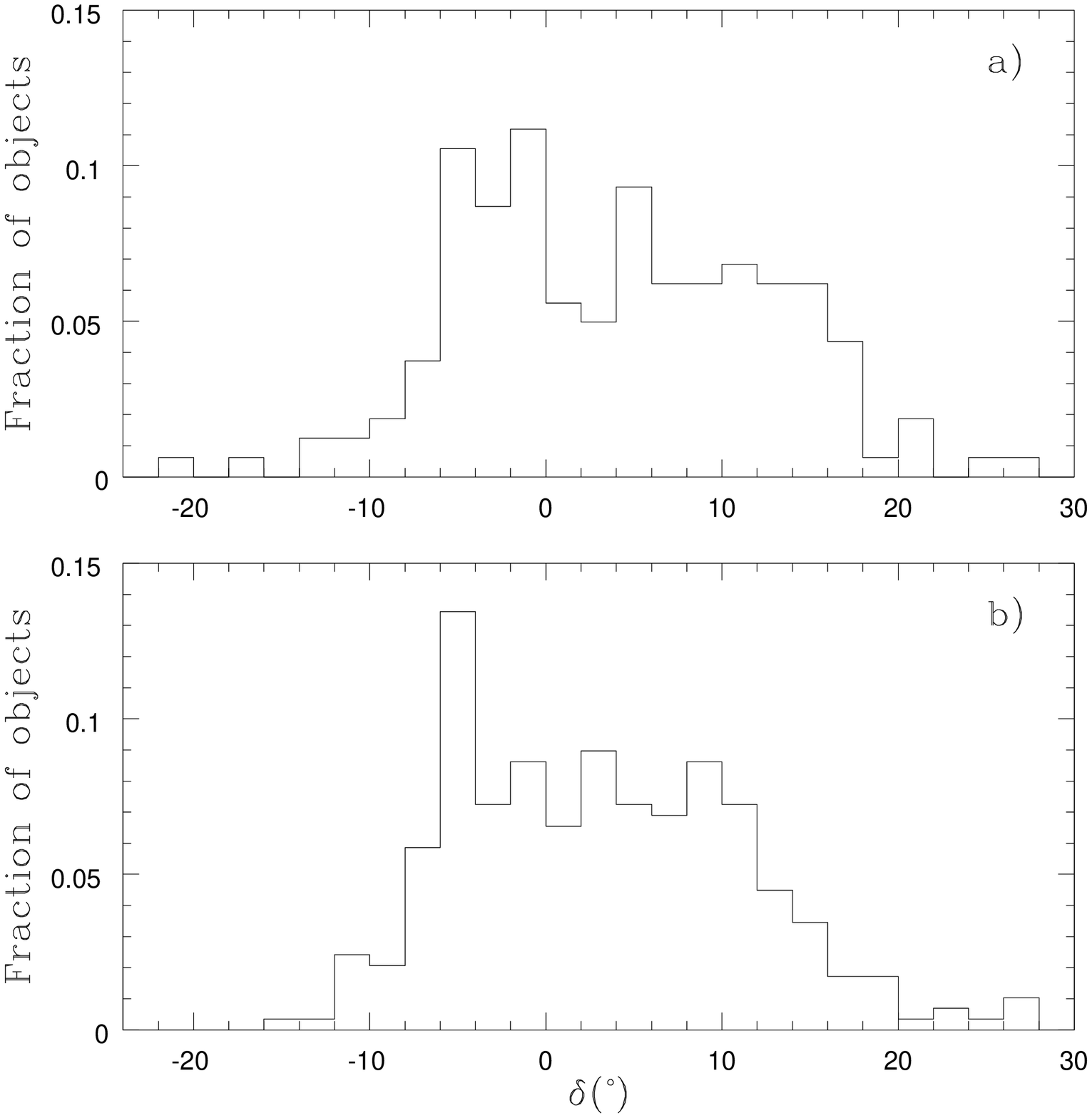} 
\epsfxsize=\hsize 
\epsfbox{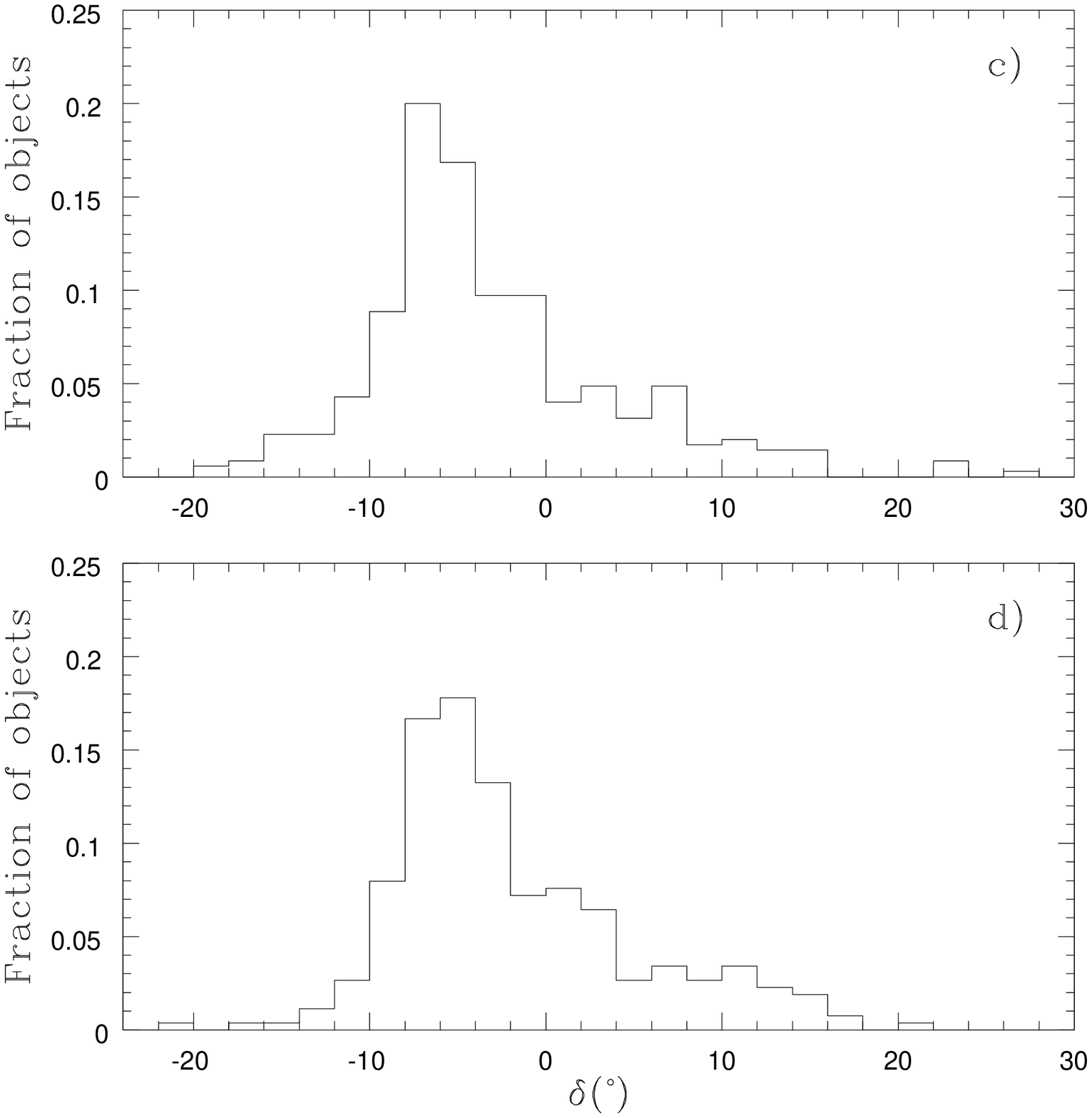}  
\caption[]{ Histograms of the $\delta$ distribution for a) the 
intercluster sample, b) the ESP sample, c) the A3528 complex,
d) the A3558 complex.  }
\label{histo_delta}
\end{figure}

The first step is the study of the morphological mix in the various samples
described in Sect.~\ref{thedata}, which represent different environments.
In order to maximize the statistics, we have divided the galaxies in three
broad morphological classes (early, intermediate and late types), defined
by ranges in $\delta$ and $(U-B)$ values.
\\ 
The first interval, which refers to early-type galaxies, is relative to values 
of $\delta\le-4^\circ$, i.e. to the region of the spectral sequence of E/S0 
Kennicutt's galaxies. The corresponding color limit (for galaxies in the
A3558 complex) is $(U-B)\ge 0.35$. The second interval is defined by  
$-4^\circ < \delta\le6^\circ$ and $0 \le (U-B) < 0.35$, and corresponds to
early spirals, from Sa to Sbc. The last interval includes the late part of 
the spectral sequence, from Sbc to irregular galaxies, and is defined by  
$\delta>6^\circ$ and $(U-B) < 0$.  
\\
\begin{figure}
\epsfysize=8.5cm
\epsfxsize=\hsize
\epsfbox{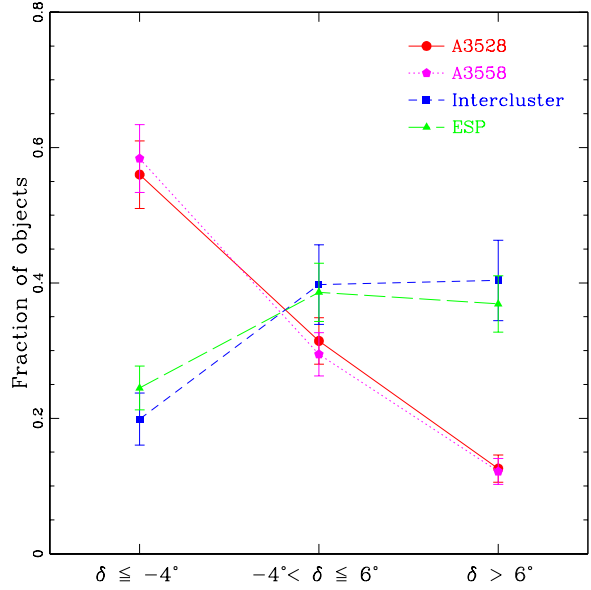}
\caption[]{ Fractions of galaxies divided in three morphological classes, for 
each spectral sample: A3528 (circles), A3558 (pentagons), intercluster 
(squares) and ESP (triangles). The vertical bars represent the 1-$\sigma$ 
statistical uncertainties. }
\label{frazioni}
\end{figure}
In Figure \ref{frazioni} the fractions of galaxies in the three morphological
bins are shown for each sample. 
As expected, cluster galaxies show a completely different morphological
distribution with respect to ESP and intercluster galaxies, being 
the fraction of early-type galaxies dominant in such high density environment.
\\
More interesting, this figure shows a broad agreement between the intercluster
and the ESP sample: this fact suggests that the galaxies located in the 
intercluster regions of the Shapley supercluster have a morphological 
mix not different from field galaxies. This agreement is confirmed also by
a more detailed analysis of the histograms of the $\delta$ distributions
(Figures \ref{histo_delta}a and \ref{histo_delta}b): 
applying a K-S test, the probability that 
they have been extracted from the same distribution is more than 40\%. 
Therefore, although the mean overdensity of intercluster galaxies is of the 
order of $(N / \bar{N})\sim 4$ (Bardelli et al. 2000), their morphological
mix has not changed significantly from that corresponding to lower density 
field galaxies (i.e. ($N / \bar{N})\sim 1$).  
\\
For what concerns cluster galaxies, there is a complete agreement between the
A3528 and A3558 complex morphological mix (Figure \ref{frazioni}). 
Small differences in the left part of the histograms in Figures 
\ref{histo_delta}c 
and \ref{histo_delta}d are due to the incompleteness of the A3558 complex 
survey: literature redshift data (present in the Figure \ref{frazioni} through 
the use of $(U-B)$ colors but not in the histograms) mainly covers the central 
region of the A3558 cluster, rich of early-type galaxies. As a consequence of
this incompleteness, the $\delta$ distributions of the two complexes are
not consistent with the hypothesis of having been extracted from the same 
distribution: however, limiting the analysis to $\delta > -11^o$, the K-S
probability rises to $\sim 20\%$. 
\\
For what concerns the $\theta$ parameter, it is clear from Figure \ref{delthe}
that a degeneracy is present along the sequence:
two galaxy spectra of very different type (for instance an elliptical and
an Sc spiral) can have the same value of $\theta$. 
Given the fact that this parameter is mainly related to the emission line
strength, we have analysed the $\theta$ distributions only for late-type 
galaxies ($\delta>6^\circ$).
Applying a K-S test to the $\theta$ distributions in each sample, we find that 
they are all consistent each other. Therefore, although the morphological 
mix in the cluster environment is significantly different from that in the 
field, it seems that the spectral properties of the late-type galaxy 
population do not change in the different environments. 

%
\subsection{Spectral morphology vs. local density }

\begin{figure}
\epsfysize=8.5cm
\epsfxsize=\hsize
\epsfbox{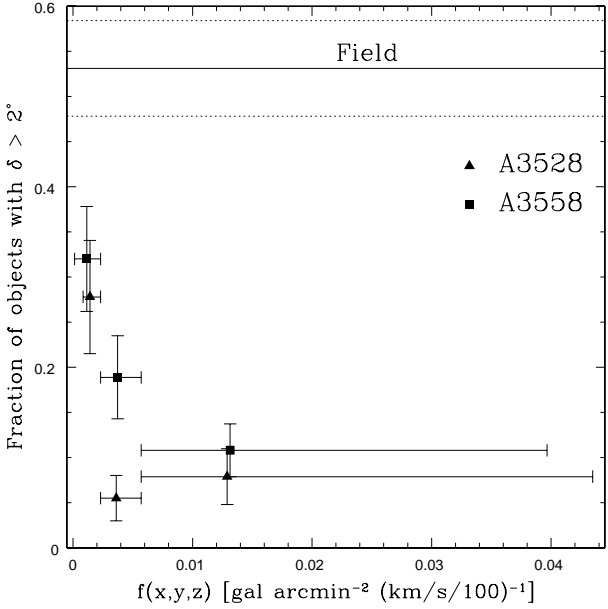}
\caption[]{ Relation between the fraction of late-type galaxies and local 
density $f(x,y,z)$ in the A3528 (triangles) and A3558 (squares) complexes,
in the magnitude range $17.0 \le b_J \le 18.8$.  
Horizontal bars indicate the extension of the density intervals:
the points are located at the position of the density average value within
the bin. Vertical bars represent the 1-$\sigma$ uncertainties. 
The reference value for the field (from ESP galaxies) is reported as a solid
line: dotted lines represent its 1-$\sigma$ uncertainties. }
\label{morfodens}
\end{figure}

A more detailed analysis of environmental effects on galaxy morphology can be
performed relating the local density with the spectral type.
A reliable estimate of the density field is possible only for the cluster 
complexes: in fact, the sampling of the intercluster region is too sparse
and non-uniform to allow a precise description of the local density point
by point. For what concerns the ESP galaxies, we consider them as a reference
sample with local density consistent with the mean density of the Universe.
\\
For the cluster complexes we have adopted the three-dimensional densities 
estimated in Bardelli 
et al. (1998b, 2001) using the DEDICA algorithm (Pisani 1993, 1996).
These densities are in units of galaxies per arcmin$^2$ per km/s/100:
given the fact that the two structures are at different redshifts 
(\vmed$_{A3558} \sim 14600$ km/s and \vmed$_{A3528} \sim 16400$ km/s), it is 
necessary to rescale the densities in order to allow a meaningful comparison
of the samples.
We chose to rescale all the values to the A3528 complex, multiplying the
A3558 densities by a factor (\vmed$_{A3528}/$\vmed$_{A3558})^2 \sim 1.3$.  
\\
For what concerns the A3528 complex, the dynamical and substructure analysis
performed in Bardelli et al. (2001) revealed the presence of a number of
clumps at $v\sim 20000$ km/s which are not dynamically part of the main 
structure, mainly associated to the presence of the A3535 cluster.
Therefore, we decided to eliminate galaxies with $v\ge 19000$
km/s in the A3528 complex analysis. 
\\
We have divided the density range in three bins, roughly equally 
populated: within these intervals we have calculated the fraction of 
galaxies with $\delta>2^\circ$ and $(U-B)<0.2$ (for literature spectra in
the A3558 complex).
This value of $\delta$ is located near the boundary between Sa and Sb galaxies
(see Figure \ref{delthe}) and therefore we measure the variations with
local density of the fraction of intermediate and late spirals and irregulars.
The results are shown in Figure \ref{morfodens}.
\\
A decreasing of the late-type galaxies fraction with increasing values
of density is clearly visible, both in the A3528 and in the A3558 complexes,
reflecting the well-known morphology--density relation.   
Moreover, for every density value, these fractions remain significantly below 
the mean field value ($53.1 \pm 5.3 \%$), derived from ESP galaxies and 
reported as a solid horizontal line in Figure \ref{morfodens}. 
Note that the mean value from intercluster galaxies is very similar 
($54.7 \pm 7.3 \%$).
\\
For what concerns galaxies with $v\ge19000$ km/s in the A3528 complex, mainly 
belonging to the poor cluster A3535, there is not a dependence of the 
morphological mix on the local density: 
the fraction of late-type spectra remains almost
constant (from lower to higher densities: 50\%, 48\% and 52\%) and with
values in agreement with those found for the field population.
This means that although the densities are rather high, the dynamical 
mechanisms have not been able to modify the population. This may suggest
that A3535 is a young structure, perhaps formed as a consequence 
of the tidal forces generated by the merging in the A3528 complex. 

Looking in more detail at Figure \ref{morfodens} we note that,
although the A3528 and the A3558 complexes show a general agreement,
the fraction of late-type galaxies at intermediate density is significantly
different in the two samples.  
The A3558 complex distribution appears more regular and smoothly decreasing,
while for the A3528 complex there is a sudden drop from the first to the
second bin: at intermediate densities the fraction of late-type galaxies
is even lower than at high densities (although within the errors).
This lack of late-type objects (or, correspondingly, excess of early-type
galaxies) is not a consequence of the chosen magnitude bin.  
Indeed, using the whole sample ($b_J \le 19.5$) we find a morphology-density
relation well consistent with the results of Figure \ref{morfodens}.
\\
Looking at the spatial distribution of intermediate density early-type 
galaxies, we find that the excess in the A3528 cluster complex is mainly
located in the Southern part, in the A3530-A3532 cluster pair. 

Finally, we checked the existence of the morphology-density relation in the
complexes considering only galaxies in the regions between clusters, where
possibly the shock fronts are located.
Although with large error bars, the trend found in these regions is consistent
with the global one: taken at face value, this fact seems to be in contradiction
with the results of Bardelli et al. (1998b), who found an excess of 
blue galaxies in the region between A3562 and A3558. 
However this discrepancy can be understood considering the limited spectral
range of our spectra ($\lambda \ga 3700$ \AA), which do not allow to fully 
sample the $U$ band. Looking in detail the spectra of the objects 
responsible of the blue excess, we find that they are classified as early-type
galaxies: the fact that these objects have $(U-B) < 0.3$ implies that
their spectra have a significant rise in the ultraviolet region (which is
outside our spectral range). 

%
\section{Equivalent width of [OII] and Star Formation Rate}
 
Other classical properties which are often used to study the
galaxy characteristics as a function of the environment are the equivalent
width of the [OII]$\lambda 3727$ emission line and the star formation rate. 
Among the various estimators for the star formation rate we adopted the formula 
of Kennicutt (1992b), as reported by Balogh et al. (1998) 
\begin{equation}
SFR= 6.75\times 10^{-12} {{ L} \over {L_{\odot}}} 
     EQW({\rm [OII]})\ \ M_{\odot}/yr
\label{eq:sfr}
\end{equation}      
where $\displaystyle{{ L} \over {L_{\odot}}}$ is the galaxy B-band luminosity 
normalized to the solar value and \eqw~is the equivalent width of the 
[OII] line (in \AA). 
The exact value of the constant, related to the dust absorption, is not 
relevant in this context because we want to make a comparison between the star 
formation rates of field and cluster galaxies placed at the same distance.

%
\subsection{Equivalent width of [OII] }

We measured the [OII] equivalent width for each galaxy fitting a Gaussian to
the [OII] line on the spectrum normalized to its continuum
(fitted with a spline3 function). Given the typical signal to noise ratio
of our spectra the detection of a line implies an equivalent width larger 
than about 5 \AA.
This procedure is the same which was applied to ESP spectra (Zucca et al.
1997).

Analysing the distribution of [OII] equivalent widths in the various samples, 
we found that the percentage of galaxies with \eqw$>5$ \AA~is, as
expected, different for field and cluster galaxies, ranging from 
44\% for the ESP sample and 57\% for the intercluster sample, to
14\% for the A3528 complex. For the A3558 complex this percentage rises to
31\%, but this higher value is a consequence of the fact that the literature
redshift data (for which spectra are not available) are mainly concentrated in 
the core of A3558, rich of early-type galaxies (see also Sect. 4.1): missing 
the equivalent width measurement of these objects, the fraction of
emission line galaxies is artificially increased.  

\begin{figure}
\epsfysize=8.5cm
\epsfxsize=\hsize
\epsfbox{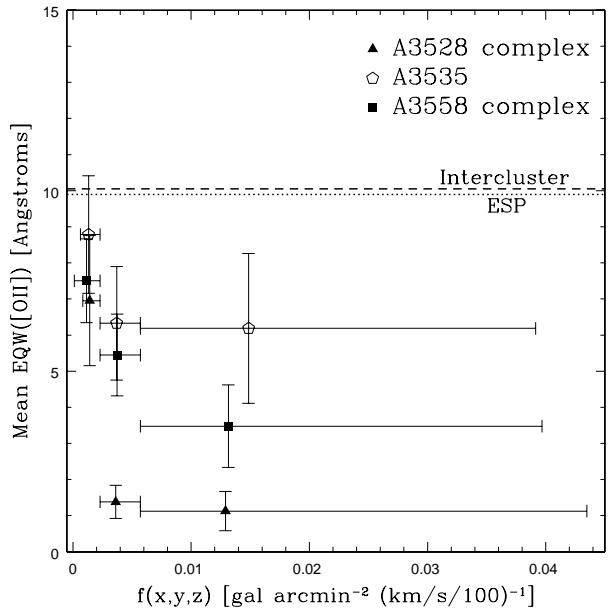}
\caption[]{ Relation between the mean [OII] equivalent width and the local 
density $f(x,y,z)$ in the A3528 (triangles) and A3558 (squares) complexes,
in the magnitude range $17.0 \le b_J \le 18.8$. Open symbols refer to the
cluster A3535.  
Horizontal bars indicate the extension of the density intervals:
the points are located at the position of the density average value within
the bin. Vertical bars represent the 1-$\sigma$ uncertainties. 
The reference values for the field are reported as a dotted line (for ESP
galaxies) and as a dashed line (for intercluster galaxies). }
\label{eqwdens}
\end{figure}

In Figure \ref{eqwdens} we show the mean rest frame \eqw~in the various
samples as a function of the local density: as a comparison, the field values
are shown as horizontal lines.
Note that the mean values were derived considering the total samples,
including also galaxies with \eqw=0. 
A clear decrease of the mean \eqw~as the density increases is visible
in both cluster complexes, although the values for the A3558 sample are
always higher. This fact can be, at least in part, a consequence of the 
missing spectra of early-type galaxies (see above).
Also the cluster A3535 follows a similar trend, but with higher values.
Moreover, all these samples are always well below the field: note also 
that no differences are present between ESP and intercluster galaxies. 

We can compare our \eqw~values with those given by Balogh et al.
(1997), who analyzed a sample of 15 rich clusters from the CNOC1 survey,
with redshift in the range $0.2 - 0.55$ and $M_R < -18.5$.  
These authors give the \eqw~values as a function of the distance from
the cluster center: for innermost cluster members ($R<0.3R_{200}$) 
\eqwmed$=0.3\pm 0.4$ \AA, for outer cluster members 
($0.3R_{200}<R<2R_{200}$) \eqwmed$=3.5\pm 0.4$ \AA, where $R_{200} \sim
1.25$ \hmpc.  
Using these limits also for our cluster galaxies, in the same luminosity range 
of Balogh et al., we find \eqwmed$=0.82\pm 0.61$ \AA~and 
$3.28\pm 0.59$ \AA~for the innermost and outer members, respectively.
The very good agreement between our estimates (which refer to low redshift
clusters) and those of CNOC1 means that there is no significant evolution
in the \eqwmed~value up to $z\sim 0.5$. 

The dependence of \eqwmed~on the local density can be due to a real
intrinsic variation of the line strength with the environment, but can be
also simply induced by the existence of a morphology-density relation.
In fact, there is a significant correlation between \eqw~and
the classification parameter $\delta$: the highest is the value of 
$\delta$ for a galaxy, the highest is its line strength. 

In order to better investigate this point, we restrict our analysis to
late-type galaxies (i.e. with $\delta>2^\circ$): 
in this case, we find that there
are no significant variations of \eqwmed~with the local density, for
all samples. The mean values, in the whole density range, are: 
\eqwmed$_{A3528}=12.71 \pm 1.96$ \AA, \eqwmed$_{A3558}=17.64 \pm 2.02$ \AA, 
\eqwmed$_{A3535}=11.87 \pm 1.76$ \AA, \eqwmed$_{ESP}=16.63 \pm 1.32$ \AA~and 
\eqwmed$_{INTERCL}=15.88 \pm 1.58$ \AA.
\\
All these values are consistent each other within $2\sigma$ and in some
cases already within $1\sigma$: therefore, it seems that the environment 
has not a strong effect on the line strength of late-type galaxies.
This fact confirms the result found using the $\theta$ distribution 
of late-type galaxies (see Sect. 5.1). 

However, to better understand the role of the environment on the galaxy
properties we need to take into account also their luminosities, and
therefore to analyse their star formation rate. 

\begin{figure}
\epsfysize=8.5cm
\epsfxsize=\hsize
\epsfbox{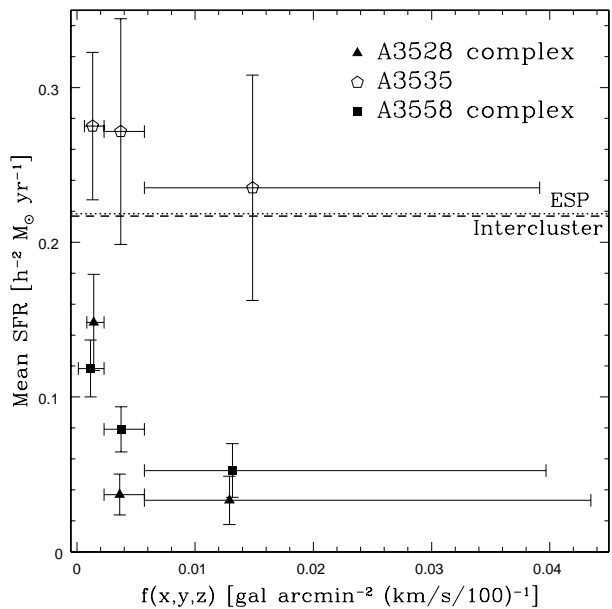}
\caption[]{ Relation between the mean star formation rate and the local 
density $f(x,y,z)$ in the A3528 (triangles) and A3558 (squares) complexes,
in the magnitude range $17.0 \le b_J \le 18.8$. Open symbols refer to the
cluster A3535.   
Horizontal bars indicate the extension of the density intervals:
the points are located at the position of the density average value within
the bin. Vertical bars represent the 1-$\sigma$ uncertainties. 
The reference values for the field are reported as a dotted line (for ESP
galaxies) and as a dashed line (for intercluster galaxies). }
\label{sfrdens}
\end{figure}

%
\subsection{Star Formation Rate}
 
We computed the star formation rate for each galaxy with measured \eqw, 
following eq.(\ref{eq:sfr}), after having corrected our $b_J$ magnitudes
for galactic absorption and having transformed them to the B-band. 

In Figure \ref{sfrdens} we show the mean star formation rate in the various
samples as a function of the local density: as a comparison, the field values
are shown as horizontal lines. Again the mean values are derived using also
galaxies with $SFR=0$. 
Also in this case there is a clear decrease of the star formation rate with
increasing densities, with values well below the field ones. The decrease
has a $3.3\sigma$ and $2.6\sigma$ significance level for the A3528 and A3558
samples, respectively. 
The cluster A3535 is the only exception, being its \sfrmed~higher than the 
field one and almost independent of the local density. 

However, as noted before for the \eqwmed, this behaviour can be induced by 
the existence of a morphology-density relation: in order to check this point 
we extrapolated the expected star formation rate for each sample following 
the method of Balogh et al. (1998). 
\\
First we derived from the field sample the mean star formation rate of 
early-type ($\langle SFR \rangle _{early}$) and late-type 
($\langle SFR \rangle _{late}$) galaxies and for each sample we computed in
each density bin the fraction of early-type ($f_{early}$) and late-type 
($f_{late}$) galaxies.
The expected mean star formation rate is then obtained as 
\begin{equation}
\langle SFR \rangle = f_{early} \langle SFR \rangle _{early} +
      f_{late} \langle SFR \rangle _{late}  
\end{equation}
Note that Balogh et al. (1998) use a photometric classification (bulge or
disk dominated) for their galaxies, while we apply a spectroscopic 
classification ($\delta>2^\circ$). 

We find that for the cluster complexes the extrapolated \sfrmed~is always
higher than the real one, of a factor $\sim 2$ for the A3528 complex
and $\sim 1.5$ for the A3558 complex. 
This trend is qualitatively similar to the results of Balogh et al. (1998),
but it has a very low statistical significance ($\sim 1 \sigma$): in fact,
these authors found a difference of a factor $\sim 5$ between the 
extrapolated and the real values of the star formation rate.

For what concerns the cluster A3535, we find an opposite behaviour: the
extrapolated \sfrmed~is lower than a factor $\sim 0.8$ with respect to the
real one, although these two values are consistent each other at $\sim 1 \sigma$
level. This fact indicates that in this particular cluster the
star formation rate has not been depressed. 

%
\section{Discussion and Summary }

In this paper we studied the morphological properties of galaxies in the
central part of the Shapley Concentration, covering an extremely wide range
of densities, from the rich cluster cores to the supercluster environment
[$(N / \bar{N}) \sim 4$]. Moreover, these results
were compared with galaxies from the ESP survey, which represent the 
reference value for the ``true" field [$(N/ \bar{N}) \sim 1$].

Given the fact that we are using galaxies at the same distance and in a well
defined magnitude range, we are not biased by evolutionary effects and by
the sampling of different parts of the luminosity function. Moreover, 
all the spectra in our survey were taken with the same telescope and
instrumental set up and therefore our sample is highly uniform.
All these characteristics allowed us to perform an accurate spectral 
classification based on a Principal Components Analysis technique. 

We found that all spectra can be well reconstructed using only three components,
representing a) an early-type galaxy spectrum, b) a spectrum with very 
prominent blue continuum and emission lines, and c) an emission line dominated 
spectrum. This classification can be parametrized using two parameters:
$\delta$, which represents the contribution of the blue over the red part
of the spectrum, and $\theta$, which represents the importance of the 
emission lines.
In the $\delta - \theta$ plane the galaxies follow a well defined sequence,
and it results that $\delta$ is enough to discriminate among the various
morphological types.

This spectral classification, together with the [OII] equivalent widths and the
star formation rates, has been used to study the properties of galaxies  
at different densities: cluster, intercluster (i.e. galaxies in the 
supercluster but outside clusters) and field environment.  

The first result is that no significant differences are present between 
samples at low density regimes (i.e. intercluster and field galaxies):
the morphological mix, the \eqwmed~and the \sfrmed~have consistent values
in these two environments. This result, although not completely unexpected,
has been proved here for the first time. 

Cluster galaxies not only have values significantly different from the field
ones, but also show a dependence on the local density.
A well defined morphology-density relation is present in the cluster complexes:
the existence of this relation is not obvious a priori in merging clusters,
because the dynamical phenomena could have shuffled galaxies from different
local densities. 
\\
A possible explanation of the persistence of the relation after a major merging
(as in the case of the A3558 complex) is that the shuffling regarded only the
peripherical galaxies (i.e. in low density regions), while the cores
have potential wells deep enough to retain the original galaxies 
and to be not influenced by this phenomenon.

Also the \eqwmed~shows a trend with the local environment, decreasing at
increasing densities.
This trend can be probably induced by the existence of a morphology-density
relation: in fact, if we consider only late-type galaxies, we find no 
significant dependence of \eqwmed~on the local density. Moreover, both field
and cluster samples have values consistent each other (within the statistical
uncertainties): therefore it seems that the environment has not a strong 
effect on the line strength of late-type galaxies.

Finally we analyzed the \sfrmed~as a function of the density, finding again
a decreasing trend. In order to verify if also this trend is a consequence
of the morphology-density relation, we extrapolated the star formation rate
expected on the basis of the field \sfrmed~values, taking into account the
morphological mix of the clusters [see Sect. 6.2 and eq.(\ref{eq:sfr})].
The extrapolated values are higher than a factor $\sim 1.5 - 2$ than the
observed ones: this result (although with low statistical significance) is
consistent with the claim of Balogh et al. (1998) that the star formation
rate in clusters is depressed.

Note that the cluster A3535, which is in the background of the A3528 complex,
deviates from these trends: it is spiral rich, with a morphological mix 
similar to the field values, and shows a \sfrmed~consistent with the 
extrapolated one. This indicates that in this particular cluster the
star formation rate has not been depressed. 

In this paper we investigated the properties of the galaxy population
in a particular, extreme environment: 
a significant improvement of these studies will be achieved by wide area 
spectroscopic survey, like 2dFGRS and SDSS, which will allow to homogeneously
sample a variety of environments, with high statistics.

%
\section*{Acknowledgements}
The authors thank B. Marano, G. Zamorani and G. Vettolani for discussions.
AB acknowledges the hospitality of the Osservatorio Astronomico di Bologna.
We also thank the referee (M.Drinkwater) for useful comments which improved 
the presentation of the results.
\\
This work has been partially supported by the Italian Space Agency grant
ASI-I-R-105-00. 

%

%


\begin{thebibliography}{99}

\bibitem{} Balogh L.M., Morris S.M., Yee H.K.C., Carlberg R.G., Ellingson E., 
           1997, ApJL 488, L75

\bibitem{} Balogh L.M., Schade D., Morris S.M., Yee H.K.C., Carlberg
           R.G., Ellingson E., 1998, ApJL 504, L75

\bibitem{} Bardelli S., Zucca E., Vettolani G., Zamorani G., 
           Scaramella R., Collins C.A., MacGillivray H.T., 1994,
           MNRAS 267, 665 

\bibitem{} Bardelli S., Zucca E., Zamorani G., Vettolani G., 
           Scaramella R., 1998a, MNRAS 296, 599

\bibitem{} Bardelli S., Pisani A., Ramella M., Zucca E., Zamorani G., 
           1998b, MNRAS 300, 589 

\bibitem{} Bardelli S., Zucca E., Zamorani G., Moscardini L., Scaramella
           R., 2000, MNRAS 312, 540

\bibitem{} Bardelli S., Zucca E., Baldi A., 2001, MNRAS 320, 387  

\bibitem{} Bellenger R., Dreux M., Felenbok P., Fernandez A., Guerin J.,
           Schmidt R., Avila G., D'Odorico S., Eckert W., Rupprecht G., 1991,
           The Messenger 65, 54

\bibitem{} Binggeli B., Sandage A., Tammann G.A., 1988, ARA\&A 26, 509

\bibitem{} Bothun G.D., 1982, PASP 94, 744

\bibitem{} Butcher H., Oemler A., 1984, ApJ 285, 426

\bibitem{} Connolly A.J., Szalay A.S., Bershady M.A., Kinney A.L., Calzetti D.,
           1995, AJ 110, 1071

\bibitem{} Dressler A., 1980, ApJ 236, 351 

\bibitem{} Felenbok P., Guerin J., Fernandez A., Cayatte V., Balkowski C.,
           Kraan-Korteweg R.C., 1997, Experimental Astronomy 7, 65

\bibitem{} Folkes S., Lahav O., Maddox S., 1996, MNRAS 283, 651

\bibitem{} Fukugita M., Shimasaku K., Ichikawa T., 1995, PASP 107, 945

\bibitem{} Galaz G., de Lapparent V., 1998, A\&A 332, 459

\bibitem{} Hoffman G.L., Lewis B.M., Salpeter E.E., 1995 ApJ 441, 28

\bibitem{} Hubble E., Humason M.L., 1931 ApJ 74, 43

\bibitem{} Hubble E., 1936 The Realm of the Nebulae (Oxford, Oxford University
           Press)

\bibitem{} Kendall M., 1980, Multivariate Analysis, Griffin, London (2nd 
           edition)

\bibitem{} Kennicutt R.C., 1992a, ApJS 79, 255

\bibitem{} Kennicutt R.C., 1992b, ApJ 388, 310

\bibitem{} Lund G., 1986, OPTOPUS - ESO Operating Manual No. 6 

\bibitem{} Metcalfe N., Godwin J.G., Peach J.V., 1994, MNRAS 267, 431

\bibitem{} Murtagh F., Heck A., 1987, Multivariate Data Analysis, Reidel

\bibitem{} O'Connell R.W., Marcum P., 1997, in Tanvir N.R., 
           Arag\'on-Salamanca A., Wall J.V. eds., The Hubble Space Telescope
	   and the high redshift Universe, p. 63

\bibitem{} Postman M., Geller M.J., Huchra J.P., 1984, ApJ 281, 95
 
\bibitem{} Pisani A., 1993, MNRAS 265, 706

\bibitem{} Pisani A., 1996, MNRAS 278, 697 

\bibitem{} Ramella M., Geller M.J., Huchra J.P., 1992, ApJ 384, 396

\bibitem{} Sandage A., 1975, in Sandage A., Sandage M., Kristian J. eds. 
           (Chicago: Univ. Chicago Press), Galaxies and the Universe, 1

\bibitem{} Sodr\'e L.Jr., Cuevas H., 1997, MNRAS 287, 137  

\bibitem{} Sodr\'e L.Jr., Cuevas H., Capelato H.V., 1998, in Wide Field Surveys
           in Cosmology, Colombi S., Mellier Y. eds, Editions Frontieres,
	   p. 424

\bibitem{} de Theije P.A.M., Katgert P., 1999, A\&A 341, 371

\bibitem{} de Vaucouleurs G., de Vaucouleurs A., 1961, Mem. R. Astron. Soc. 68,
           69

\bibitem{} Yentis D.J., Cruddace R.G., Gursky H., Stuart B.V., Wallin J.F.,
           MacGillivray H.T., Collins C.A., 1992, in MacGillivray H.T.,
           Collins C.A. eds, Digitized Optical Sky Surveys, Kluwer, 
           Dordrecht, p.67  

\bibitem{} Vettolani G., Zucca E., Zamorani G. et al., 1997, A\&A 325, 954

\bibitem{} Vettolani G., Zucca E., Merighi R. et al., 1998, A\&AS 130, 323

\bibitem{} Zaritsky D., Zabludoff A.I., Willick J.A., 1995, AJ 110, 1602

\bibitem{} Zucca E., Zamorani G., Vettolani G. et al., 1997, A\&A 326, 477

\end{thebibliography}
\end{document}